# Near-unity quantum yield from carbon nanotube excitons coupled to

# plasmonic nanocavities

*Yue Luo[1], Ehsaneh D. Ahmadi[1], Kamran Shayan[1], Yichen Ma[1], Kevin S. Mistry[2],*

*Changjian Zhang[3], James Hone[3], Jeffrey L. Blackburn[2], and Stefan Strauf [1*]*

[1] Department of Physics, Stevens Institute of Technology, Hoboken, NJ 07030, USA

[2] National Renewable Energy Laboratory, Golden, Colorado 80401, United States

[3] Department of Mechanical Engineering, Columbia University, New York, NY 10027, USA

*Address correspondence to: strauf@stevens.edu

**Single-walled carbon nanotubes (SWCNTs) are promising absorbers and emitters to enable novel photonic and optoelectronic applications but are also known to severely suffer from low optical quantum yields. Here we demonstrate SWCNTs excitons coupled to plasmonic nanocavities reaching deeply into the Purcell regime with $F_P$=234 (average $F_P$=76), near unity quantum yields of 70% (average 41%), and a photon emission rate of 1.7 MHz into the first lens. The measured ultra-narrow exciton linewidth (18 μeV) implies furthermore generation of indistinguishable single photons from a SWCNT. To demonstrate utility beyond quantum light sources we show that nanocavity-coupled SWCNTs perform as single-molecule thermometers detecting plasmonically induced heat (ΔT=150K) in a unique interplay of excitons, phonons, and plasmons at the nanoscale.**

**Introduction**

Single-walled carbon nanotubes (SWCNTs) are promising absorbers and emitters in applications including solar energy conversion,[1] biological imaging[2], and on-chip quantum photonics such as room-temperature single photon emission in oxygen-doped SWCNTs[3] and quantum optical circuits with integrated superconducting detectors[4]. Optical quantum yields from SWCNTs remain nevertheless low (2-7%)[5–7] and light enhancement strategies are required for practical implementations to improve quantum yield[8], preserve exciton coherence[9], and potentially enable indistinguishable photons[10]. Early attempts to couple SWCNT excitons to planar surface plasmons show however no Purcell effect[11], while dielectric cavities display only moderate Purcell factors of $F_P=5$ (Ref.7).

A promising alternative to the dielectric cavity is the plasmonic nanocavity that features nanometer sized gaps with ultra-small mode volumes resulting in drastically enhanced spontaneous emission (SE) rates of quantum emitters[8,12–14]. Plasmonics can be lossy when dipole emitters are placed in close proximity to isolated metal nanoparticles leading to dominant nonradiative (NR) recombination, i.e. photoluminescence (PL) quenching[15]. This, however, is not the case for nanoplasmonic resonators, which can feature a dominant radiative recombination rate with a branching ratio of 75% radiative to 25% NR when emitters are in close proximity (2 nm) to nanocavity gap-modes[16]. Importantly, the higher-order dark modes in isolated plasmonic nanoparticles become hybrid-plasmonic dipole modes in both bowtie nanoantennas and gap-mode structures, resulting in massively enhanced emitter decay via radiative channels[17]. This effect recently led to the first experimental demonstration of a single molecule entering the strong-coupling regime with a plasmonic gap-mode[18]. Remarkable results have also been demonstrated for dye molecules coupled to elliptical gold nanoparticles with intensity enhancement factors (EF) of about 100 (Ref.19), single molecule emission from antenna-in-a-box platforms (EF~1100)[20], and bowtie nanoantennas (EF~1340)[21]. Using plasmonic gap modes, large Purcell factors up to $F_P=1000$ were also demonstrated[8,16]. Plasmonic nanocavities are thus an appealing route to overcome the low quantum yield of carbon nanotubes and to enable efficiency advance in device applications.

Here we demonstrate SWCNTs excitons coupled to plasmonic nanocavities reaching deeply into the Purcell regime with $F_P=234$, near unity quantum yields of 70%, and a photon emission rate of 1.7 MHz into the first lens. The ultra-narrow exciton linewidth (18 μeV) implies generation of indistinguishable single photons from a SWCNT. In addition, we show that nanocavity-coupled SWCNTs perform as single-molecule thermometers detecting plasmonically induced heat ($\Delta T=150K$) in a unique interplay of excitons, phonons, and plasmons at the nanoscale.

## Results

### Design and characterization of plasmonic nanocavities

With the goal of combining the superior optical properties of co-polymer wrapped SWCNTs featuring long exciton coherence times[22] and the unmatched light enhancement properties of plasmonic nanocavities featuring gap-modes, we designed dense dimer nanogap antenna arrays where the SWCNT is located on top (**Fig. 1a**). To guide the structural design of the plasmonic chips we carried out finite-difference time domain (FDTD) simulations to achieve spectral matching of emitter and mode (Supplementary Information **Fig.S1**). As shown in **Figure 1b**, the intensity enhancement provided by the gap mode can be three orders of magnitude. However, the measured EF of the exciton emission can be significantly lower due to spatial mismatch between emitter and mode, lateral orientation mismatch of 1D SWCNTs, variations of the exciton emission energy with nanotube chirality, short circuit of the plasmon mode through residual SWCNT conductivity, and PL quenching due to direct contact with the metal.

Several techniques allow us to systematically overcome these challenges, resulting in efficient emitter mode coupling. First of all the bowtie antenna arrays, fabricated via electron beam lithography (**Fig. 1c-e**), where covered with a 2 nm thick $Al_2O_3$ spacer layer. This provides the required distance from the gap mode to achieve dominant radiative recombination[16] and effectively prevents PL quenching and short circuiting of the plasmon. In addition, the dense plasmonic array with an antenna spacing of 500 nm provides a large light collection efficiency enhancement of $\varepsilon=5.1$ (66%) compared to off-chip areas (13%), that is maintained even for large spatial detuning (Supplementary Information **Fig.S2**). By varying the geometry parameters of the nanocavity one can achieve spectral resonance which is rather straightforward since the typical Q-factors of plasmonic modes are rather low (Q=2-20)[23], enabling the simultaneous coupling of exciton absorption and emission dipoles to the cavity mode, as demonstrated experimentally in **Fig. 1f.** To drastically enhance the chance that individual semiconducting SWCNT with known chirality and emission wavelength are spectrally coupled we have synthesized small-diameter SWCNTs via laser-vaporization. Co-polymer (PFO-BPy) wrapping of these SWCNTs[5] produces dispersions with large numbers of (5,4), and (6,4) semiconducting SWCNTs with emission wavelengths well-positioned for coupling to the cavity resonance. The SWCNT dispersions fabricated in this way were dried out on the plasmonic chips.

### Spatial mapping and quantum light emission of individual SWCNTs

**Figure 2a** shows a 2D spatial scan of stray light off of the sample surface revealing the outline of the markers and bowtie arrays, while it shows in addition large regions of enhanced scattered light corresponding to the deposited molecular material. To clearly identify SWCNTs, hyperspectral images were recorded for the Raman G band signal, which is found regardless of tube chirality. The corresponding 2D scan in **Fig. 2b** reveals bright spots corresponding to individual SWCNTs at a moderate density in about half of the area. To clearly address the subset of either (5,4) or (6,4) SWCNTs we have carried out hyperspectral PL mapping in **Figs. 2c,d** covering the corresponding exciton PL emission bands located at 855 nm and 880 nm, respectively. Individual SWCNTs appear as nearly spherical dots as expected, since the SWCNT length (0.5-1 μm) is close to the excitation laser spot size. In this way, individual SWCNTs with known chirality can be addressed for further investigations with respect to their coupling strength to the plasmonic mode (blue box in **Fig. 2d**).

To create a strong Purcell effect, the lateral orientation of the SWCNT must match both the pump polarization for efficient light absorption as well as the plasmon mode polarization. Simulations show that when the incident polarization of the pump laser is parallel to the long axis of the bowtie structure (p-pol), the field enhancement is 2 orders higher as compared to the orthogonal direction (s-pol), a condition easy to fulfill since the orientation of the bowties is known from imaging (**Fig. 1c-e**). To probe for the subset of SWCNTs that are oriented along the bowtie dimer axis we recorded hyperspectral maps for s-pol and p-pol excitation (**Fig. 3**). While some SWCNTs are found that display only a weak intensity variation and are thus either spatially away from the hot-spot or orientationally misaligned, others clearly switch off with s-pol excitation. Several SWCNTs are found displaying a rather pronounced PL polarization extinction contrast with one example of 25:1 following the plasmon mode orientation within a few degrees (**Fig. 3b**). This strong PL extinction ratio is in contrast to SWCNTs that are located on bare wafers with values varying from 3:1 to 5:1 [24]. The significantly stronger polarization contrast demonstrated here is a direct indication of exciton-plasmon mode coupling. In addition, **Fig. 3c** demonstrates pronounced photon antibunching signatures of the nanocavity coupled exciton emission with a second-order coherence function of $g^2(\tau=0)=0.30\pm0.06$ when recorded with a 10 nm broad bandpass filter and normalized to the average peak area beyond the first two side peaks that are affected by bunching. Further improvements to the single-photon purity could be achieved by narrowband filtering, e.g. with a spectrometer [7]. The observed values are nevertheless already significantly below $g^2(\tau=0)=0.5$ and thus clearly prove the quantum-dot like (0D) nature of the exciton emission [3,24], while the bunching signature is comparable to our findings for SWCNTs in dielectric cavities in the presence of blinking at the ns time scale [3,24].

**Determination of plasmonic enhancement factor**

While the strong polarization contrast is a good indicator for exciton-plasmon coupling the quantitative strength is best characterized by the plasmonic enhancement factor EF. In the saturation regime EF has contributions from three processes, an enhanced absorption rate α, enhanced light extraction ε, and enhanced SE emission rate $\gamma_R/\gamma_{R0}$, such that EF = $\alpha\varepsilon\gamma_R/\gamma_{R0}$ [8]. We define "coupled SWCNTs" as those spectrally matched with the plasmon mode and located on the bowtie arrays, but not necessarily fully laterally or orientationally aligned to the high field area of the mode. In contrast, we define "uncoupled SWCNTs" as those that are found far from the 100×100 μm patterned areas. For recording light-light curves from the uncoupled SWCNTs we have in every case aligned the pump laser polarization with the nanotube axis to maximize emission. For the coupled SWCNTs the laser polarization was fixed along the bowtie long axis to efficiently excite the gap-mode that is also polarized along this axis. **Fig. 4a** displays the integrated intensity versus pump power comparing uncoupled (blue dots) and coupled SWCNTs (red dots) that increases linearly until slight saturation sets in at highest pump powers. This behavior can be modeled with a three-level rate equation analysis of 0D quantum-dot like excitons (Supplementary Information **Fig.S3**). Apparently, every on-chip SWCNT displays significantly higher intensity compared to every off-chip SWCNTs, indicating that virtually all emitters on the plasmonic array are coupled. Statistical analysis shows an average EF= 22±3, a best case of EF=90, and even EF=7 for the worst case (Supplementary Information **Fig.S4d**). This omnipresent enhancement is expected since the average SWCNT length (0.5-1 μm) is larger than the nanoantenna spacing of 400 nm, and since that the light collection efficiency varies only slightly from ε=5.1 for perfect positioning down to ε=3.5 for large spatial detuning of 80 nm. As a result, the measured photon emission rate from the SWCNT reaches up to 120000 cts/s, which after correcting for the detection system efficiency of 7±2% yields 1.7 million cts/s emitted into the first lens.

**Determination of Purcell factor and cavity-enhanced quantum yield**

The finding of EF=90 implies that a significant contribution of the measured enhancement stems from the Purcell effect of the underlying radiative rate $\gamma_{R0}$. Using the calculated ε=5.1, and α=3.2±0.2 as determined from the linewidth broadening in **Fig.5a** (vide infra), one can determine the SE rate enhancement $\gamma_R/\gamma_{R0}$ =5.5±0.2 (550%) in this way. To directly measure the SE lifetime $\tau=1/\gamma$ of the $E_{11}$ exciton emission we carried out time-correlated single photon counting (TCSPC) measurements shown in **Fig.4b.** The experimental data have been deconvolved using the system response function and yield mono-exponential decay time fits for this SWCNT pair of $\tau_{uncoupled} = 248 \pm 3$ ps and $\tau_{coupled} = 37\pm3$ ps, respectively. Statistical analysis of 20 SWCNTs shown in **Fig.4c-d** reveals that all uncoupled SWCNTs display decay times within a narrow range with an average value of $\tau_{coupled}$=215 ps and a variance of 20 ps. In contrast, every on-chip SWCNT displays a faster rate that varies from 155±3 ps to 37±3 ps in the best case (see Supplementary Information **Fig.S4**). The total decay rate enhancement $\gamma_{tot}$, defined as the ratio $\tau_{uncoupled}$ / $\tau_{coupled}$ that includes the contribution from NR recombination is plotted in **Fig.4d** together with the corresponding values of the intensity enhancement EF. The clear correlation demonstrates that faster measured decay results directly in more light emitted from the SWCNT, implying that the measured rate enhancement are dominated by radiative decay. For the best case the TCSPC experiment yields a six-fold total rate enhancement, i.e. $\gamma_{tot}$=6.7±0.5 (670%). We note that comparing $\gamma_R/\gamma_{R0}$ =550%, which contains only radiative rates in EF, to $\gamma_{tot}$ from the measured decay rates that contain also the NR contribution allows determining the branching ratio. This comparison reveals that radiative recombination contributes 82% (NR recombination 18%) to the measured emission decay, matching previous findings for emitters in close proximity (2 nm) to nanoplasmonics gap-modes[14].

Such a high SE rate enhancement is quite remarkable, particularly in light of recent reports of PFO-wrapped SWCNTs coupled to high-Q dielectric cavities in the Purcell regime that show measured SE rate enhancement of only 10% [7]. It is important to note that SWCNTs suffer in general from rather low radiative quantum yield $\eta = \frac{\tau_{R0}^{-1}}{\tau_{R0}^{-1}+\tau_{NR}^{-1}}$ of 2-7 % [5–7], where $\tau_{R0}$ and $\tau_{NR}$ are the radiative and non-radiative decay times of the uncoupled system. A precise estimate of $\eta = 2 \pm 0.5\%$ for PFO-wrapped SWCNTs was recently determined from single-photon emission saturation[7], i.e. a 50-fold faster NR rate compared to the radiative rate in the absence of a Purcell effect. Assuming our PFO-wrapped SWCNTs also have $\eta = 2\%$ for the uncoupled case, one can calculate the underlying Purcell factor from the relations:

$$\tau^{-1}_{\text{coupled}} = \frac{1+F_p}{\tau_{R0}} + \frac{1}{\tau_{NR}} + \frac{1}{\tau_M} \quad \text{and} \quad \tau^{-1}_{\text{uncoupled}} = \frac{1}{\tau_{R0}} + \frac{1}{\tau_{NR}} \ ,$$

where $1/\tau_M$ is the metal loss rate. In this way, Jeantet *et al.*[7] determine an underlying $F_P$=5±2 for the measured 10% rate enhancement. Including losses for our system through the branching ratio (82%:18%), the data in **Fig.4b** yield $F_p = 234$ for the best case with an average of $F_p$ =76 and a minimum of $F_p$ =11 (Supplementary Information **Fig.S4c**). While this is by far the highest Purcell factor ever reported for the exciton emission from SWCNTs it underperforms the theoretical value ($F_P$ =6591) 28-fold, predominantly due to orientational and/or spatial mismatch, leaving room for future improvements using deterministic assembly. Such a high $F_P$ implies that the SE coupling factor β=$F_P$/(1+$F_P$) is unity (99.6%), indicating that virtually all of the SE emission is coupled to the nanocavity mode[25]. In addition, the cavity-modified quantum yield of the SWCNT can now be estimated to be $\eta^{cav} = 70\%$ for the best case with an average of 41% as illustrated in **Fig.4e**. A key result of the pronounced Purcell effect up to $F_P$=234 is that the initially low quantum yield of 2% approaches almost unity due to the coupling to the plasmonic gap-mode that enhances the radiative rate to become ultimately faster than the NR losses in the system.

**Ultra-narrow linewidth regime**

The demonstrated large light enhancement and improved internal quantum efficiency enables one to study the exciton photophysics in two extreme regimes – at ultra-low optical pump powers where pump-induced exciton dephasing is minimized as well as at high pump powers where plasmonic heating is expected. Specifically, we have studied the zero-phonon linewidth (ZPL) of the $E_{11}$ exciton emission over five orders of magnitude variation in excitation pump power (**Fig.5a**). For the uncoupled SWCNT a ZPL of 70 µeV is found at 10 µW pump power. This is almost twice better than the previous record for PFO-SWCNT material[22] and is a result of the 2 nm thick $Al_2O_3$ layer that strongly reduces spectral diffusion typically found when SWCNTs touch $SiO_2$ or glass substrates[24]. At lowest pump powers of 50 nW the plasmonically coupled SWCNT displays an ultra-narrow ZPL of only $18 \pm 3$ µeV shown in **Fig.5a** (see Supplementary Information for deconvolution fits). This linewidth is the narrowest ever reported for carbon nanotube excitons, i.e. 7-fold narrower than previous best reports for PFO-wrapped SWCNTs[22] and twice narrower than the resolution-limited linewidth reported for surfactant-free air-suspended SWCNTs[26].

**Plasmonic heating regime - single-molecule thermometry**

At the other extreme of high pump powers, we demonstrate that plasmonically induced heating can be measured in the near-field based on the peculiar changes in the exciton spectrum of an individual PFO-wrapped SWCNT. With increasing pump power the acoustic-phonon wings that peak about 2 meV symmetrically around the ZPL strongly gain in intensity until they fully merge with the ZPL (**Fig.5b**). In total, the pump induced broadening and plasmonic heating contribute to a 560-fold spectral broadening of the exciton emission from Γ=18 µeV to Γ=10 meV, when the pump power is raised from 50 nW to 2 mW. In contrast, the uncoupled SWCNT is unaffected by the plasmonic heating and saturates at about Γ=200 µeV at highest pump powers (**Fig.5a**). Similar spectral evolution can be observed by exciting the coupled SWCNT only moderately (200 µW), but with progressive increase of the base temperature, as shown in **Fig. 5c.** It is thus clear that the spectral changes accompanying high power (>300 µW) excitation of the coupled SWCNT are associated with plasmonically induced heating. Our detailed exciton lineshape analysis in Ref. 22 (involving the temperature dependent boson occupation factor, see Supplementary Information) demonstrated that this spectral evolution is caused by the thermal break-up of the acoustic-phonon confinement provided by the co-polymer backbone of the PFO-wrapped SWCNTs. This model allows us to extract the plasmonically induced rise in temperature at the single-nanotube level. The extracted "single-molecule read-out" matches well with thermal transport modelling of the total heat power confined on the bowtie antenna, reaching 150 K at 2 mW optical pump power (**Fig.5d**). Perfect spatial matching for thermal coupling is apparently easier to achieve as compared to the case of Purcell enhancement, which is also evident from the heat map in **Fig.5d** showing that the entire area of the bowtie antenna (250 nm side length) contributes to heating. We note that unlike previous demonstrations of measuring heat in plasmonic nanostructures,[27] we directly determine here the thermal near-field through an individual molecule rather than a distributed ensemble in a thin film. The demonstrated utility of PFO-wrapped SWCNTs as single-molecule thermometers with all-optical read out uncovers the unique interplay of excitons, phonons, and plasmons at the nanoscale. One can envision that this approach is of use for the development efforts of photothermal cancer therapy and diagnostics that are based on nanoplasmonics.

## Discussion

Our results reveal a new way to characterize the enhanced absorption rate in plasmonically coupled systems. **Fig. 5a** shows that at any given pump power P the coupled SWCNT has a larger ZPL in comparison to the uncoupled SWCNT by a factor 1.75-1.86. The additional ZPL broadening can be attributed to an effectively higher excitation power due to the enhanced absorption provided by the plasmonic mode. The magnitude can be quantified since the coupled SWCNT in **Fig.5a** displays a clear trend over 3 orders of magnitude in P that fits to a square root function, i.e. $\Gamma \sim \sqrt{P}$. Using this functional dependence one can estimate a cavity enhanced absorption from the linewidth data to result in $\alpha$=3.2±0.2 for this case. We note that while pump-induced broadening in 1D SWCNTs is often attributed to exciton-exciton scattering (EES)[28], our observation of strong antibunching precludes EES as the dominant contribution to the ZPL since the number of excitons is well below 2 ($g^2(\tau=0)\leq0.5$) even at P=200 µW (**Fig.3c**). Instead, the observed $\sqrt{P}$ dependence of $\Gamma$ reflects the behavior of 0D excitons in quantum dots that are governed by motional narrowing effects of fluctuating charges in the exciton vicinity[29]. This is also in agreement with the observed side-peak blinking in the $g^2(\tau)$ trace **Fig.3c** revealing fluctuations at the ns time scale[25]. Our analysis shows that the pump power-induced dephasing is a useful resource to quantitatively determine enhanced absorption rates in cavity-coupled systems.

Using the well-known relation $\mathrm{FWHM} = 2\Gamma = \frac{2\hbar}{T_2} = \frac{\hbar}{T_1} + \frac{2\hbar}{T_2^*}$ , where FWHM is the full with at half maximum of the ZPL shown in **Fig. 5a**, one can estimate the $T_2$ coherence time of the exciton emission to be at least 73 ps. Together with the radiative SE time $T_1$ = 45 ps, that can be estimated from the measured decay rate $\tau_{coupled}$ =37±3 ps and the radiative to NR branching ratio, one can estimate $T_2/2T_1$ = 0.81, implying that dephasing is almost purely defined by radiative decay and emitted single photons are indistinguishable[10]. This ratio directly characterizes the two-photon interference (TPI) visibility $V=T_2/2T_1$ as can be measured with a Hong-Ou-Mandel type experiment. As shown in the inset in **Fig.5a**, for higher pump powers, pump-induced dephasing starts to dominate the contribution to the ZPL, leading to reduction of the $T_2$ time and thus the TPI visibility. Allowing $V=1/e=0.37$, which occurs at a pump power of 1 μW, a single photon emission rate of 1100 cts/s into the fist lens can be determined, making such a measurement a priori feasible with low dark count APD detectors (see Supplementary Information). Albeit only at lowest pump powers, it is quite remarkable that indistinguishable single photons can be generated from SWCNTs excitons coupled to nanocavities deeply in the Purcell regime.

The advances demonstrated in our study, particularly on improved radiative efficiency, have immediate implications for a number of SWCNT-based applications, such as deep-tissue in vivo fluorescence imaging in the near-infrared region as well as on-chip quantum information processing. The advent of indistinguishable single photons from a carbon nanotube, which was here achieved by the interplay of decoupling acoustic phonons to prolong exciton coherence together with the large SE Purcell enhancement provided by the nanocavity, could enable advanced quantum information protocols on-chip that go beyond single photons, such as entanglement protocols implemented via post selection. However, to become practical, the emitted rate of indistinguishable photons needs to be significantly improved, which could be achieved by protecting excitons against pump-induced dephasing when trapped in the energetically deep localization potentials of solitary dopant atoms[3]. While perhaps less obvious, enhancing radiative efficiency also has a direct impact on improving the power conversion efficiency of solar energy harvesting devices (including photovoltaics and solar fuels devices), due to the generalized optoelectronic reciprocity theorem[30]. In particular, non-radiative decay reduces the maximum attainable open-circuit voltage for a photovoltaic solar cell. Thus, any solar energy harvesting application benefits directly from approaches that provide either better fundamental understanding of, or technological strategies to improve, the ultimate radiative efficiency of a given semiconductor.

**Methods**

**Plasmonic chip fabrication:** The bowtie arrays were fabricated at the CUNY-ASRC facility by electron-beam lithography (EBL) using 1:3 diluted Methyl Styrene/ Chloromethyl Acrylate Copolymer (ZEP520A) in anisole that was spin-coated at 2500 rpm onto the Si/SiO2 substrate. The side lengths of the individual bowtie antennas triangles were 250 nm, with a height of 30 nm and gap size varying from 10-20 nm. The samples were subsequently patterned in an Elionix ELS-G100 EBL system and developed at a chilled temperature of 2°C in n-Amyl acetate for 120 sec. To convert the polymer template into a plasmonic array we deposited a 3 nm Cr adhesion layer and 30 nm Au metal on the samples in an electron beam evaporator (AJA Orion 3-TH) followed by liftoff in Dimethylacetamide solvent at room temperature. Finally, a 2nm thick layer of $Al_2O_3$ was deposited by atomic layer deposition.

**Carbon nanotube synthesis and dispersion:** SWCNTs with the natural isotope ratio (99% $^{12}$C) were synthesized by the laser vaporization (LV) process, as described previously[31]. The small-diameter SWCNTs synthesized for this study were produced at a furnace temperature of 800 °C in the LV process, and all syntheses were run at a power density of ~100 W/cm$^2$ ($\lambda$ = 1064 nm, Nd:YAG). The target consisted of Alfa Aesar graphite (2-15 μm, stock #14736), and 3 wt % each nickel and cobalt catalysts. SWCNTs grown in this way were dispersed in poly[(9,9-dioctylfluorenyl-2,7-diyl)-alt-co-(6,60-{2,20-bipyridine})] (PFO-BPy). Briefly, 1 mg/mL of raw LV soot was mixed into a solution of 2 mg/mL of PFO-BPy in toluene. This solution was then sonicated with a 1/2 in. probe tip for 30 min at 40% power (Cole-Parmer CPX 750) in a bath of cool (18°C) flowing water for heat dissipation. After sonication, solutions were centrifuged at 30 000g for 5 min using a SW32-Ti rotor (Beckman). Finally, PFO-BPy dispersed LV SWCNTs in toluene were deposited directly onto the array region on the sample followed by 105 °C baking on hot plate for 3 hours.

**Photoluminescence spectroscopy:** Micro-photoluminescence (μ-PL) measurements were taken inside a closed-cycle cryogen-free cryostat with a 3.8 K base temperature and ultralow vibration (attodry1100). Samples were excited with a laser diode operating at 780 nm in continuous wave mode. A laser spot size of about 0.85 micron was achieved using a cryogenic microscope objective with numerical aperture of 0.82. The relative position between sample and laser spot was adjusted with cryogenic piezo-electric xyz-stepper while 2D scan images were recorded with a cryogenic 2D-piezo scanner (attocube). Spectral emission from the sample was collected in a multimode fiber, dispersed using a 0.75 m focal length spectrometer, and imaged by a liquid nitrogen cooled silicon CCD camera. For high resolution measurements we used an 1800-groove grating and 10 microns slit width settings. Laser stray light was rejected using an 800 nm RazorEdge ultrasteep long-pass edge filter. For time-resolved PL lifetime measurement light from a supercontinuum laser (NKT photonics) operating at 78 MHz repetition rate and 7-ps pulse width was filtered by a 10 nm bandpass filter centered at 780 nm. Time-correlated single photon counting (TCSPC) was carried out with a coincidence counter (SensL) and a fast avalanche photodiode (APD) with a timing jitter of 39 ps (IDQuantique). The system response function was measured sending scattered laser light from the sample surface with the 800 nm RazorEdge filter removed and at an APD count rate of about 10 KHz to match with the rate of the $E_{11}$ exciton emission. The second-order coherence function $g^2(\tau)$ was recorded with the same TCSPC setup but in Hanbury-Brown and Twiss configuration where the exciton emission is first sent through a 50:50 multimode fiber splitter before reaching the two APDs.

**Plasmonic resonance measurement:** To determine the transmission/scattering spectrum of the plasmonic bowtie arrays we have fabricated samples on glass substrate with equivalent dimensions to the ones on $Si/SiO_2$ substrates used for the SWCNT studies but at rather low density of bowties being spaced out at least 5 μm to avoid plasmonic lattice effects[23]. Transmission measurements were carried out under perpendicular incidence excitation with a tungsten-halogen white light source filtered by a broadband linear polarizer. Passing the polarizer the white light was focused onto the samples by a 100× microscope objective and the transmitted light through the sample was collected by a 50× microscope objective and sent through a variable spatial filter before coupling into a multi-mode fiber attached to a high-resolution grating spectrometer with TE-cooled CCD camera.

*Acknowledgement.* We like to thank Milan Begliarbekov for supporting the EBL process development at the City University of New York Advanced Science Research Center (ASRC) nanofabrication facility. S.S. and J.H. acknowledge financial support by the National Science Foundation (NSF) under award DMR-1506711. S.S. acknowledges financial support under NSF award ECCS-MRI-1531237. J.B. and K.M. gratefully acknowledge funding from the Solar Photochemistry Program of the U.S. Department of Energy, Office of Science, Basic Energy Sciences, Division of Chemical Sciences, Geosciences and Biosciences, under Contract No. DE-AC36-08GO28308 to NREL

## *Author contributions*

S.S. and Y.L. designed the experiment. Y.L., K.S. and C.Z. performed the optical experiments and/or analyzed the data. E.D.A fabricated the plasmonic chips and together with Y.M. contributed theory analysis. K.S.M. and J.L.B. have grown and dispersed SWCNT material. S.S., J.H., J.L.B. and Y.L. co-wrote the paper. All authors discussed results and commented on the manuscript.

## *Competing financial interests*
The authors declare no competing financial interests.

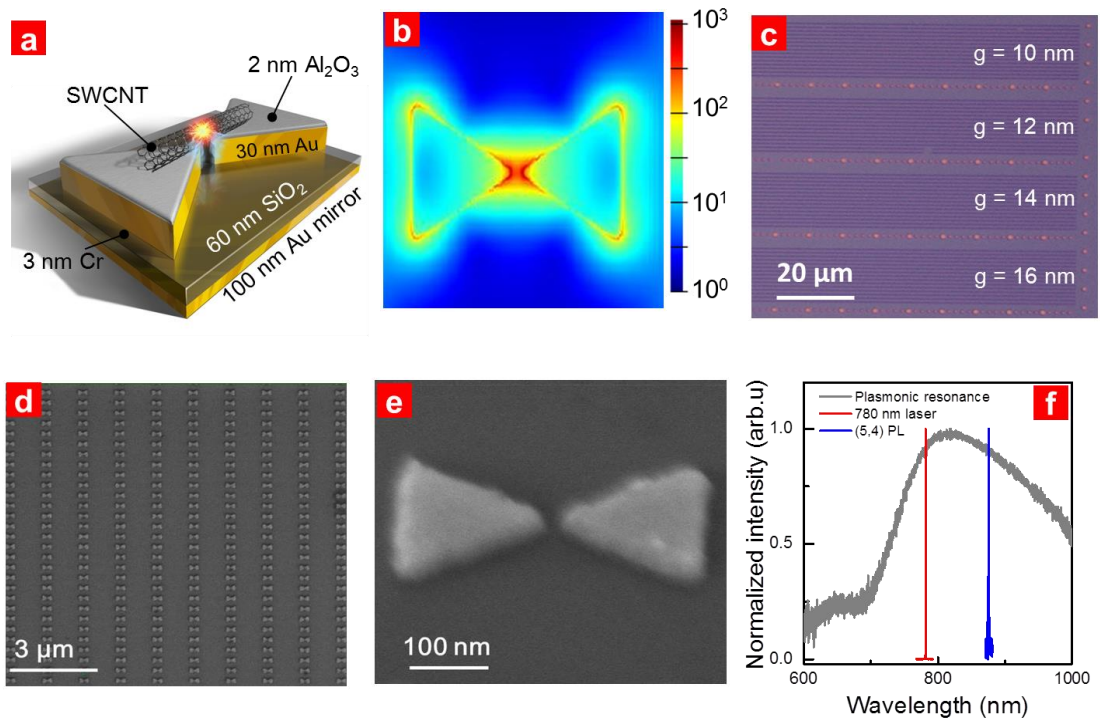

**Figure 1 Overview of plasmonically coupled carbon nanotube system. a,** Shematic of an individual SWCNT suspended across a bowtie antaennae. The SWCNT (d < 1 nm) is portrayed with significantly larger scale than actual size for clarity. The SWCNT is separated from the plasmonic gold substrate by a 2 nm ALD-grown $Al_2O_3$ spacer layer to prevent optical quenching and short circuit of the nanoplasmonic gap-mode underneath. **b,** FDTD simulation of the corresponding field enhancement distribution profile including finite apex angles with 3 nm radius. **c,** Bright-field optical microscope image of the plasmonic array showing four 20x100 µm stripes each containing bowtie antennas with fixed gap size g varying among stripes from 10-20 nm. The larger features are gold markers to enable repositioning to individual SWCNTs. **d,** The scanning electron microscope image shows high uniformity and orientation control of the plasmonic system. **e,** Zoom into an individual bowtie antenna with 10 nm gap showing sharp and straigth edges. **f,** Plasmon resonance spectrum (Q=6) recorded in dark-field transmission geometry (grey) together with 780 nm pump laser spectrum (red) and exciton emission spectrum of a (5,4) SWCNT (blue) showing spectral resonance is fullfilled simultaneously for both SWCNT absorption and emission.

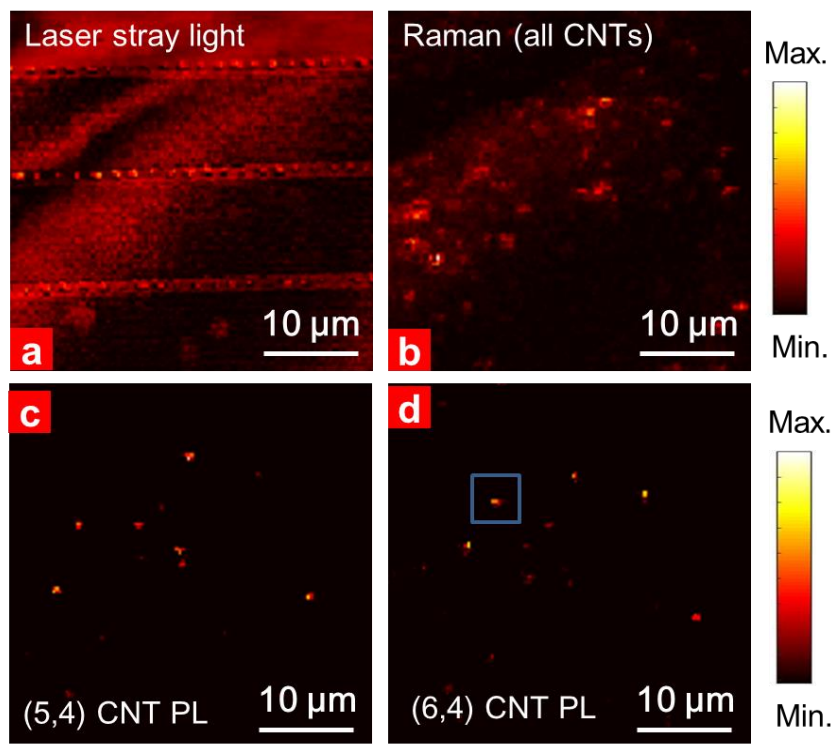

**Figure 2 Hyperspectral imaging of SWCNT distribution. a,** Laser stray light scanning off of sample surface revealing structural features similar to Figure 1c and also signatures of deposited material. **b,** Hyperspectral Raman imaging of same sample surface area, capturing the G-mode phonon of all SWCNTs. **c,** Hyperspectral PL map of same area revealing (5,4) SWCNTs filtered at 855±10 nm **d,** Hyperspectral PL map of same area revealing (6,4) SWCNTs filtered at 880±10 nm. Data are recorded at a sample temperature of 3.8 K.

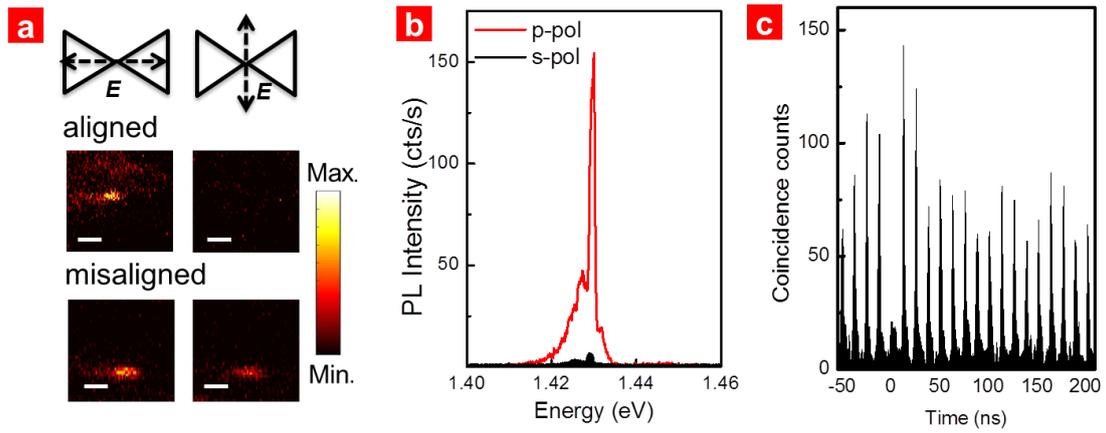

**Figure 3 Polarization dependence and quantum light signature of exciton emission. a,** Top: Schematic of excitation laser with linear polarization set to be either parallel (p-pol, left) or perpendicular (s-pol, right) to the long axis of bowtie dimer. Middle: PL maps of individual SWCNT that are well-aligned along the dimer axis for both p-pol (left) and s-pol (right) excitation. Bottom: Similarly recorded polarization-dependent PL maps for a SWCNT that is orientationally misaligned. Scale bars are 1 μm. **b,** Corresponding PL spectra for a well-aligned SWCNT featuring a large excitation polarization extinction ratio of 25:1. **c,** Second-order coherence function $g^2(\tau)$ recorded at 200 μW excitation power demonstrating pronounced single photon antibunching with $g^2(0)=0.30\pm0.06$.

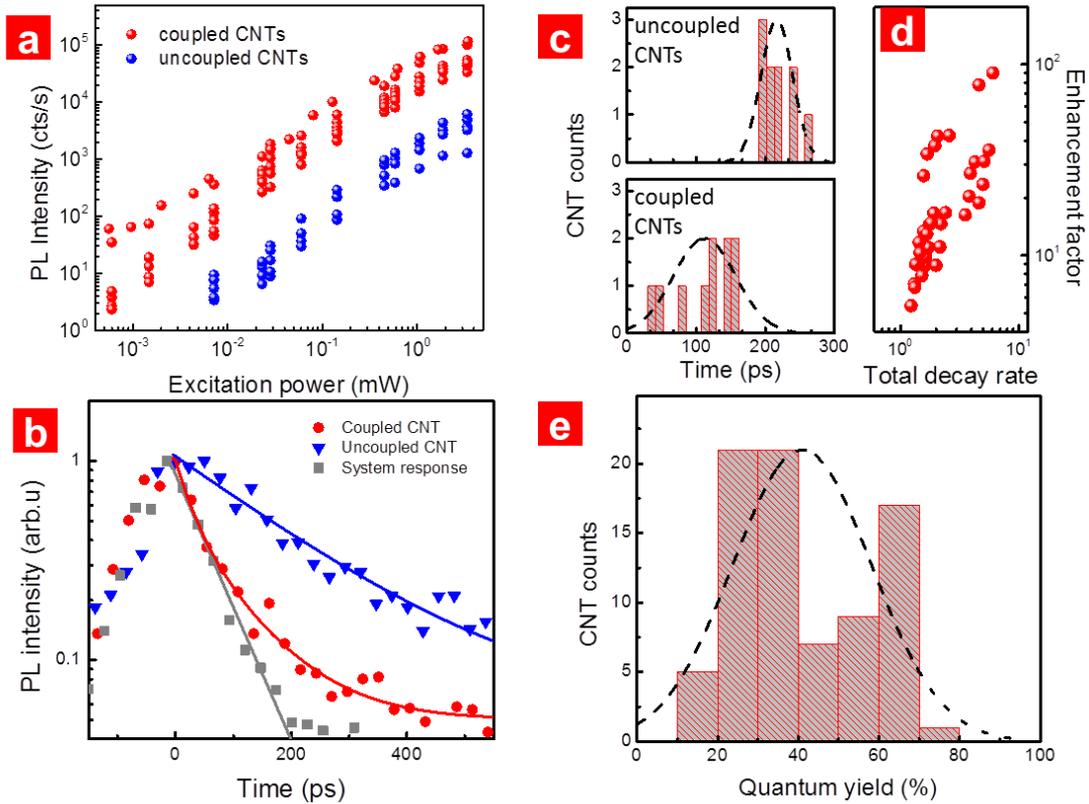

**Figure 4 Quantifying SE enhancement of the E$_{11}$ exciton emission. a,** Integrated PL intensity of the E$_{11}$ zero-phonon line as a function of excitation power. Red circles are data for coupled SWCNTs and blue triangles are from uncoupled (off-chip) SWCNTs. The corresponding enhancement factor EF determined at highest pump powers displays an average value of EF=22 with a best case of EF=90. **b,** Time dynamics of E$_{11}$ exciton emission recorded by TCSPC at 200 μW excitation power. Grey squares: System response for back-reflected laser light. Solid grey line: Mono-exponential fit representing the system response. Blue triangles are data for an uncoupled SWCNT and the solid blue line is a deconvolved fit which yields a mono-exponential decay time of τ$_{uncoupled}$=248±3 ps. The red circles are data from a coupled SWCNT and yield τ$_{coupled}$=37±3 ps (red solid line). **c,** Corresponding histogram of lifetimes for 10 uncoupled SWCNTs (upper panel) and 10 coupled SWCNTs (lower panel). **d,** Correlation between total decay rate enhancement and intensity enhancement factor EF. **e,** Histogram of estimated plasmonically enhanced quantum yield for 81 SWCNT-pairs showing η$^{cav}$=70% for the best case and 41% for the average where the distribution function (dashed line) has its maximum. All data are recorded at 3.8 K.

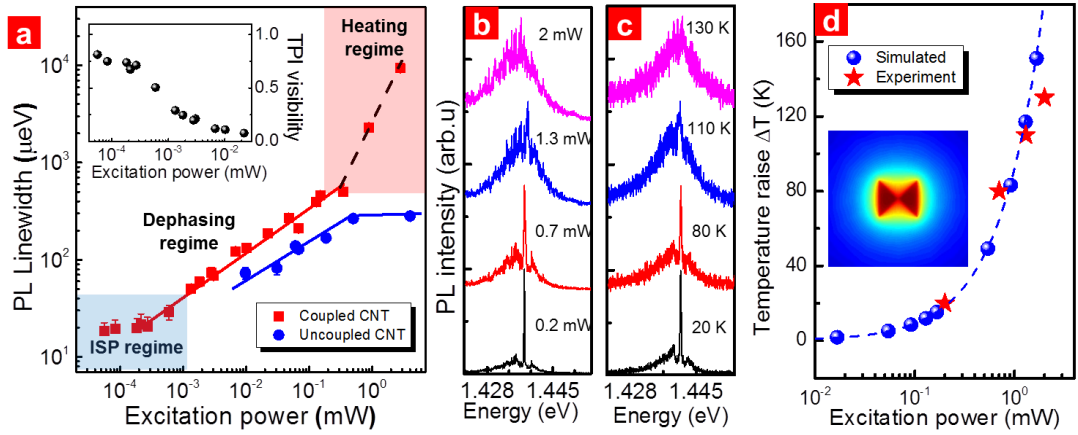

**Figure 5 Excitation power and temperature dependence of the E₁₁ exciton emission linewidth. a,** Spectral linewidth as a function of laser excitation power (780 nm) for a plasmonically coupled SWCNT (red squares) and uncoupled SWCNT (blue circles). The blue box highlights the regime of indistinguishable single photon (ISP) emission where $T_2/2T_1 \geq 1/e$ (0.37). Inset: Two-photon interference (TPI) visibility versus exciation power. Data are recorded at 3.8 K. **b,** PL spectra for the coupled SWCNT recorded in the plasmonic heating regime where a significant lattice temperature increase causes the break-up of the acoutic-phonon confinement leading to strong linewidth broadening. **c,** Comparision spectra for the coupled SWCNT recorded under moderate pump powers of 200 μW and for increasing temperature from 20 K to 130 K. **d,** FDTD simuation of temperature raise ΔT of the bowtie structure as a function of excitation power (blue dots). Inset: Simulated heat map of the bowtie antenna. The red stars are the data points from the experiements in panel b with the temperature equivalent taken from the experiment in panel c.